\documentstyle[epsfig,12pt,preprint,tighten,aps]{revtex}
\begin{document}

\rightline{\large October 2002}

\vskip 2.3cm

\centerline{\huge Are four neutrino models ruled out?}
\vskip 2.4cm
\centerline{\Large R. Foot}
\vskip 0.7cm
\centerline{foot@physics.unimelb.edu.au}
\centerline{\large \it School of Physics}
\centerline{\large \it University of Melbourne}
\centerline{\large \it Victoria 3010 Australia}

\vskip 3cm
We show explicitly that four neutrino models of 
the 2+2 variety still provide
an acceptable global fit to the solar, atmospheric and LSND neutrino data.
The goodness of fit, defined in the usual way,
is found to be 0.26 for the simplest such model.
That is, we find that there is a 26\% probability of
obtaining a worse global fit to the neutrino data
We also make some specific comments on the paper,
``Ruling out four-neutrino oscillation interpretations
of the LSND anomaly'' [hep-ph/0207157], and explain
why they reached drastically different conclusions.

\newpage

One area of excitement in physics is coming from the
direction of  various neutrino experiments. There is now a range
of evidence, from atmospheric, solar and terrestrial
experiments strongly supporting the idea
that neutrinos have mass and change flavour
via oscillations. At the present time, the precise
oscillation pattern remains uncertain and is the
subject of much speculation.

Recently, it was argued\cite{fv} that 
the totality of the
neutrino data now suggests an essentially unique picture:
\begin{eqnarray}
\nu_e &\to &\nu_\tau \ {\rm \ large \ angle \ oscillations \ explains \ 
the \ solar 
\ neutrino \ problem}
\nonumber \\
\nu_\mu &\to &\nu_s \ {\rm \ large \ angle \ oscillations \ explains \ 
the \ atmospheric \ neutrino
\ anomaly}
\nonumber \\
\bar \nu_e &\to &\bar \nu_\mu \ {\rm \ small\ angle \ oscillations \ explain 
\ the \ LSND \ data}
\label{1}
\end{eqnarray} 
where $\nu_s$ is a hypothetical (effectively) sterile neutrino.
Although this scheme is not very popular 
it at least
has the virtue that it will be tested in the near future:
MiniBooNE will test the oscillation explanation of the LSND anomaly, 
while the forthcoming long baseline experiments will discriminate between the
$\nu_{\mu} \to \nu_s$ and $\nu_{\mu} \to \nu_{\tau}$
possibilities for resolving the atmospheric neutrino anomaly.

Anyway, the aim of this brief note
is to explicitly estimate the overall goodness of fit (g.o.f) of the
above scheme [Eq.(\ref{1})]. 
We find that the g.o.f is 0.26 which is reasonable, but 
is perhaps controversial because
this value is more than 5 orders of magnitude larger than
the value found in Ref.\cite{mal}. As we will discuss,
the origin of this discrepancy is quite easy to
explain.

The oscillation scheme [Eq.(\ref{1})] is essentially
unique in the sense that it is the simplest scheme
involving only two-flavour oscillations
explaining the totality of the data and also specific
features such as SNO's neutral current/charge current
solar flux measurement\cite{sno}. 
Of course, other,
but more complicated schemes involving 
multi-flavour oscillations, 
are possible because they can also provide an 
acceptable fit to the data for
a range of parameters.
For example,
one can have an additional parameter,
$\sin^2 \omega$, where $\sin^2 \omega = 0$
corresponds to the scheme, Eq.(\ref{1}),
$\sin^2 \omega = 1$ is similar to Eq.(\ref{1}) with
$\nu_\tau$ interchanged with $\nu_s$ and
intermediate values of $\sin^2 \omega$
corresponds to mixed active+sterile oscillations\cite{other}.
Such schemes are called 2+2 models 
because they feature two pairs of almost degenerate states
separated by the LSND mass gap.
While the scheme of Eq.(\ref{1}) could be viewed
as a particular 2+2 scheme with $\sin^2 \omega = 0$,
it could alternatively be viewed as motivating the
following hypothesis\cite{fv}:
The fundamental theory of neutrino mixing, whatever it is,
features (i) large (or even maximal) $\nu_\mu \to \nu_s$
mixing, (ii) small-angle active-active mixing except for the 
$\nu_e \to \nu_\tau$ channel which is large.

Certainly the atmospheric fit of 
$\nu_\mu \to \nu_s$ oscillations is not perfect, 
having a goodness of fit of $0.055$\cite{sk}. 
Essentially there seems to be two main possibilities here.
First, it might be that
the atmospheric neutrino anomaly is due to (predominantly) $\nu_\mu \to
\nu_\tau$ oscillations
and the solar neutrino anomaly is also due to (predominantly) 
active-active oscillations in a 3-flavour scheme
called bi-maximal mixing\cite{bm}. In this case, it suggests that 
the LSND anomaly is not caused by neutrino oscillations at all.
The other possibility is the one identified above in Eq.(\ref{1}). 
While the fit to the atmospheric data is not perfect in
the scheme [Eq.(\ref{1})] it does
accomodate the LSND data and also has the
theoretical virtue of explaining the data using only two-flavour
oscillations. 
Both possibilities are possible [as we will explicitly show in the 
case of the scheme of Eq.(\ref{1})]
and only future data can compellingly distinguish between
the scenarios.

The overall g.o.f for the scheme,
Eq.(\ref{1}), can easily be evaluated. Things
are especially simple, since Eq.(\ref{1}) involves
only two-flavour oscillations. 
Basically there are three contributions to the $\chi^2$:
\begin{eqnarray}
\chi^2 (total) = \chi^2 (atm)  + \chi^2 (solar) + \chi^2 (LSND)
\end{eqnarray}
To avoid controversy, we use the super-Kamiokande
result for the atmospheric $\nu_\mu \to \nu_s$ 
hypothesis\cite{sk}.
For solar neutrinos, we use the value for $\chi^2 (min)$
obtained by Ref.\cite{hol} applicable for $\nu_e \to
\nu_\tau$ oscillations (this is a fit to both super-Kamiokande
and the latest SNO solar neutrino results). Finally,
for the LSND experiment we estimate a $\chi^2$ minimum
from Figure.16 
(taking the bins with $E$ in the range $20 < E/MeV < 52$)
and Figure.24
(taking the bins with $L/E$ in the range $0.5 \stackrel{<}{\sim} L/E \stackrel{<}{\sim} 1.2$)
of Ref.\cite{lsnd}\footnote{
These particular bins are chosen because they have more
than 5 events, so that the $\chi^2$ statistic should follow
the $\chi^2$ probability density function. Of course,
the details of how we analyse the LSND data does 
not strongly affect our conclusion.}. 
The result of these extractions
is:
\begin{eqnarray}
\chi^2 (atm) &\simeq & 222 \ for \ N_{atm} = 190 \ d.o.f. 
\nonumber \\
\chi^2 (solar) &\simeq & 65 \ for \ N_{solar} = 78 \ d.o.f. 
\nonumber \\
\chi^2 (LSND) &\approx & 4 \ for \ N_{LSND} = 8 \ d.o.f.
\end{eqnarray}
This gives a total $\chi^2 (total) \equiv \chi^2 (atm) + \chi^2 (solar)
+ \chi^2 (LSND) = 291$ for $N_{total} \equiv N_{atm}+N_{solar}+N_{LSND} 
= 276$ d.o.f.  This corresponds to a g.o.f of 0.26.
That is, there is a 26\% probability of obtaining a worse
global fit to the neutrino data used.
This explicitly demonstrates that the scheme Eq.(\ref{1})
currently provides a reasonable global 
fit to the neutrino data. 

Of course, it may also be legitimate to worry about whether
such a scheme can fit certain subsets of the data.
For example, super-Kamiokande identified 3 subsets
of data which might potentially discriminate between
the $\nu_\mu \to \nu_\tau$ and $\nu_\mu \to \nu_s$
solutions to the atmospheric neutrino anomaly\cite{sk2}. 
However, this discrimination, was not particularly
stringent (super-Kamiokande found that the $\nu_\mu \to \nu_s$  
possibility had a g.o.f of about $0.01$), and an alternative
analysis\cite{f} of the same data gave a much better fit (a 
g.o.f of about $0.1$).
Clearly, this discrimination is crucial and is
still one of the main issues that needs to be resolved.

As a final remark, let us also consider the recent analysis of
2+2 schemes by Maltoni, Schwetz, Tortola and Valle (MSTV)\cite{mal}.
In the MSTV paper, it
was found that the g.o.f of 2+2 models to the 
totality of neutrino data (solar, atmospheric and LSND)
was just $10^{-6}$, a quite remarkable result given
our estimate of 0.26 (above) for a particular 2+2 
model!
It turns out that it is very easy to identify the origin
of the discrepancy. Our result above was based on standard
g.o.f obtained using standard
$\chi^2$ statistics in the usual way, 
while the result of Ref.\cite{mal} was
based on a new type of statistics (which we
will denote as MSTV statistics).
In MSTV statistics, the value
of $\chi^2$ (min) is {\it not} compared with the number
of d.o.f to obtain a g.o.f.
Instead, the
value of $\chi^2$ (min) is compared with another $\chi^2-like$ 
quantity, obtained by 
fitting different data sets with different
values of the parameter $\sin^2 \omega$ [the parameter
describing the particular mixture of active+sterile
involved in the solar and atmospheric oscillations]
although, physically it can only take one value
within the model. Following a certain prescription, MSTV
then obtains a quantity which they call g.o.f. 
Clearly,
the MSTV g.o.f is a sort of relative thing obtained by comparing
the standard $\chi^2 (min)$ with some unphysical $\chi^2-like$
function.  Thus, the g.o.f calculated by MSTV is nothing to do with
the standard value, but some other quantity. 
[This is obvious since as we have seen, it is different to the standard
g.o.f by more than 5 orders of magnitude!].
Really, to be meaningful MSTV 
would have to show that their g.o.f has some
{\it absolute} statistical significance {\it and} why 
the standard g.o.f breaks down so badly for four neutrino
models.

In summary, we have explicitly 
demonstrated that at least some four neutrino models 
still provide a reasonable
description to the totality of the neutrino data. 
Clearly, more experimental
input is necessary to distinguish between the various oscillation schemes
still allowed by the data
before definite conclusions can be drawn. 
In particular, a clear discrimination between
the $\nu_\mu \to \nu_\tau$ and $\nu_\mu \to \nu_s$ atmospheric oscillation
hypothesis and a confirmation or otherwise of the 
LSND result by mini-Boone are required.

\vskip 1.6cm
\noindent
{\bf \large Acknowledgements}
\vskip 0.5cm
\noindent
The author wishes to thank R. R. Volkas and T. L. Yoon
for various discussions.

\end{document}